\documentclass[11pt, onecolumn]{IEEEtran}

\usepackage[dvips]{graphicx}
\usepackage[dvips]{color}
\usepackage{epsf,graphicx,amssymb,verbatim}
\usepackage[mathscr]{eucal}

\def\BibTeX{{\rmfamily B\kern-.05em{\scshape i\kern-.025em b}\kern-.08em
\TeX}}

\usepackage{color}

\newcommand{\less}{\le}
\newcommand{\greater}{\ge}
\newcommand{\fraction}{\frac}

\newcommand{\ti} {\tilde}

\newcommand{\graph}{{\cal   G}}

\newcommand{\scenario}{u}
\newcommand{\scenarios}{{\mathcal U}}
\newcommand{\decrease}{w}
\newcommand{\flows}{{\mathcal H}}
\newcommand{\flow}{{\mathcal F}}
\newcommand{\subsets}{{\mathcal G}}
\newcommand{\timeshare}{\lambda}
\newcommand{\duration}{T}
\newcommand{\subflow}{{\mathcal S}}
\newcommand{\flowsize}{f}
\newcommand{\potential}{\phi}
\newcommand{\parameter}{\alpha}
\newcommand{\score}{y}

\newcommand{\links} {{\mathcal L}}

\newcommand{\source}{s}
\newcommand{\sink}{d}
\newcommand{\capacity}{C}
\newcommand{\bcapacity}{\bar{\capacity}}
\newcommand{\transmissions}{{\mathcal P}}
\newcommand{\maxQQ}{Br_Q}
\newcommand{\maxQc}{Br_c}

\newcommand{\change}{\delta}
\newcommand{\inflow}{r}
\newcommand{\uncoded}{\nu}
\newcommand{\poison}{\pi}
\newcommand{\remedy}{\rho}
\newcommand{\coding}{\gamma}
\newcommand{\splitting}{\sigma}
\newcommand{\decoding}{\eta}

\newcommand{\limiting}{{\mathcal G}}
\newcommand{\intersections}{{\mathcal I}}
\newcommand{\weight}{w}
\newcommand{\bQ}{\bar{Q}}

\newcommand{\bU}{\bar{U}}

\newcommand{\origins}{{\mathcal O}}
\newcommand{\destinations}{{\mathcal D}}
\newcommand{\approximate}{{\ti{l}}}

\newcommand{\length}{l}

\newcommand{\nodes}{{\mathcal N}}

\newcommand{\packetsize}{p}

\newcommand{\ratio}{\rho}

\begin{document}

\title{Polynomial-time algorithms for coding across multiple unicasts}
\author{Tracey Ho}
\maketitle

\begin{abstract}
We consider the problem of network coding across multiple unicasts. We give, for wired and wireless networks,
efficient polynomial time algorithms for finding optimal network codes within the class of network codes
restricted to XOR coding between pairs of flows.
\end{abstract}

\section{Introduction}

In this paper we consider network coding across multiple unicasts, using the class of pairwise XOR codes
introduced in~\cite{ratnakar05linear} for wired networks. This class of codes includes ``reverse carpooling''
and two-flow ``star coding'' for wireless networks~\cite{effros06tiling}. We give efficient polynomial time
algorithms for finding optimal network codes within this class on wired and wireless networks.

\section{Problem setup}
\subsection{Multiple unicasts problem on wired network}

$K$ unicast sessions are transmitted over a network represented as a directed  graph $\graph =(\nodes,\links)$
of $N=|\nodes|$ nodes and $M=|\links|$ links.  Network coding is limited to XOR coding between pairs of uncoded
flows. Each session $c=1,\dots,K$, demands a communication rate $\inflow_c$. Each link $(a,b)\in\links$ has a
capacity denoted $\capacity_{ab}$, which, if greater than $\max_c\inflow_c$, is set to $\max_c\inflow_c$. The
total incoming capacity and total outgoing capacity of each node is upper bounded by $\bcapacity$. A solution
for a given set of demanded rates $\{\inflow_c\}$ is an assignment of values to variables
$$\{\uncoded_{ab}^{cv},\uncoded_{ab}^{cvi},\poison_{ib}^{\{c,c'\} j}
,\poison_{ib}^{cc' j},\remedy_{ab}^{cc'j},\coding_i^{\{cv, c'v'\}},\splitting_i^{\{c,c'\}},
\decoding_i^{cc'j}\}$$ satisfying:
\begin{eqnarray*} 
 &&\sum_b \left(\uncoded_{ib}^{cv}+\uncoded_{ib}^{cvi}\right)+ \sum_{c'\ne c,v'}\coding_i^{\{cv,
c'v'\}}=\sum_a\uncoded_{ai}^{cv} +\sum_{v'}\uncoded_{vi}^{cv'v}
+\sum_{c'}\decoding_i^{cc'v}+\left\{\begin{array}{ll} \inflow_c&i=v =\source_c\\0&v\ne \source_c,\sink_c,
i\ne\sink_c\end{array}\right.\nonumber\\& & \sum_b\poison_{ib}^{\{c,c'\} j}
+\splitting_i^{\{c,c'\}j}=\sum_a\poison_{ai}^{\{c,c'\}j} +\sum_{v,v'}\coding_i^{\{cv,c'v'\}} \nonumber\\& &
\sum_b\poison_{ib}^{cc' j}+\decoding_i^{cc'j}=\sum_a \poison_{ai}^{cc'j} +\splitting_i^{\{c,c'\}j} \nonumber\\&
& \sum_b\remedy_{ib}^{cc' j}+\decoding_i^{cc'j}=\sum_a\remedy_{ai}^{cc'j} +\sum_{v}\coding_j^{\{cv,c'i\}}
\nonumber\\& & \sum_{c ,v}(\uncoded_{ab}^{cv}+\uncoded_{ab}^{cva})+ \sum_{c,c',j}\left(\remedy_{ab}^{cc'j}
+\poison_{ab}^{cc'j} \right) +\sum_{\{c,c'\},j}\poison_{ab}^{\{c,c'\}j} \le \capacity_{ab}
\end{eqnarray*}

We show how  to find a solution for the problem with rates $\{\inflow_c\}$ if there exists a solution for the
problem with slightly higher rates $\{(1+2\epsilon)\inflow_c\}$ for any $\epsilon >0$.

We define a number  of queues at each node $i$ which can be interpreted as follows:
\begin{itemize}
\item $U_i^{cv}$: uncoded session $c$ data previously at node $v$\item $P_i^{\{c,c'\}j}$: data from sessions
$c,c'$ coded at node $j$ meant for both sinks\item $P_i^{cc'j}$: data from sessions $c,c'$ coded at node $j$
meant for sink $\sink_c$\item $R_i^{cc'j}$: remedy for session $c$ data that has been coded with $c'$ data at
node $j$
\end{itemize}

Data can be transmitted on a link $(a,b)\in\links$ from queue $U_a^{cv}$ to $U_b^{cv}$, from $U_a^{cv}$ to
$U_b^{ca}$, from $R_a^{cc'j}$ to $R_b^{cc'j}$, from $P_a^{\{ c,c'\}j}$ to $P_b^{\{ c,c'\}j}$, or from
$P_a^{cc'j}$ to $P_b^{cc'j}$. A coding operation at a node $a$ transforms $f$ units from each of a pair of
queues $(U_a^{cv},U_a^{c'v'})$ into $f$ units in each of the queues $R_{v'}^{cc'a}$ and $R_v^{c'ca}$. A decoding
operation at $a$ transforms $f$ units from each of a pair of queues $(R_a^{cc'j},P_a^{cc'j})$ into $f$ units in
$U_a^{cj}$. A branching operation at $a$ transforms $f$ units from a queue $P_a^{\{ c,c'\}j}$ into $f$ units in
each of the queues $P_a^{cc'j}$ and $P_a^{c'cj}$.

Then the solution variables can be interpreted as follows:\begin{itemize}\item $\uncoded_{ab}^{cv}$: average
flow rate from $U_a^{cv}$ to $U_b^{cv}$\item $\uncoded_{ab}^{cva}$: average flow rate from $U_a^{cv}$ to
$U_b^{ca}$\item $\poison_{ab}^{\{c,c'\} j}$: average flow rate from $P_a^{\{ c,c'\}j}$ to $P_b^{\{
c,c'\}j}$\item $\poison_{ab}^{cc' j}$: average flow rate from $P_a^{cc'j}$ to $P_b^{cc'j}$\item
$\remedy_{ab}^{cc'j}$: average flow rate from $R_a^{cc'j}$ to $R_b^{cc'j}$\item $\coding_a^{\{cv, c'v'\}}$:
average rate of coding transformation from $(U_a^{cv},U_a^{c'v'})$ to $(R_{v'}^{cc'a},R_v^{c'ca})$\item $
\decoding_a^{cc'j}$: average rate of decoding from $(R_a^{cc'j},P_a^{cc'j})$ to $U_a^{cj}$\item
$\splitting_a^{\{c,c'\}}$: average rate of branching transformation from $P_a^{\{ c,c'\}j}$ to
$(P_a^{cc'j},P_a^{c'cj})$\end{itemize}

\subsection{Modified problem}
\label{modified}

For a given problem instance, we consider a modified problem any solution of which is equivalent to a solution
of the original problem and vice versa. The modified problem reverses the direction of the poison flows and
imposes some additional constraints. Specifically, it is defined as follows:\begin{itemize}\item Each source
node $\source_c$ has a source queue $U^c$, 
an overflow queue $\bU^c$, and a virtual source link of capacity $\bcapacity$ from $U^c$ to queue
$U^{c\source_c}_{\source_c}$.\item Each node $a$ has a virtual coding link, a virtual decoding link and a
virtual branching link, each of capacity $\bcapacity/2$.

\item Each real and virtual link $e$ is associated with a set $\transmissions_{e}$ of pairs
$(\origins,\destinations)$ such that data units from each queue in the set $\origins$ are transformed into an
equivalent number of units in each queue in the set $\destinations$ via $e$:\begin{itemize}\item if $e$ is the
virtual source link for session $c$, $\transmissions_e=(U^c,U_{\source_c}^{c\source_c})$\item if $e$ is a real
link $(a,b)\in\links$,
$$\hspace{- 0.2 in}\transmissions_{e} =\{(U_a^{cv},U_b^{cv}),(U_a^{cv},U_b^{ca}),
(P_b^{\{c,c'\}j},P_a^{\{c,c'\}j}), (P_b^{cc'j},P_a^{cc'j}),(R_a^{cc'j},R_b^{cc'j})\}$$
\item if $e$ is the virtual coding link at node $a$,
$\transmissions_e=\left\{\left((U_a^{cv},U_a^{c'v'}),(R_{v'}^{cc'a},R_v^{c'ca})\right):\; c\ne c';a\ne
v,v'\right\}$\item if $e$ is the virtual decoding link at node $a$,
$\transmissions_e=\left\{\left(R_a^{cc'j},(P_a^{cc'j},U_a^{cj})\right):\; c\ne c';a\ne j\right\}$\item if $e$ is
the virtual branching link at $a$, $\transmissions_e=\left\{\left((P_a^{cc'j} ,P_a^{c'cj}),P_a^{\{
c,c'\}j}\right):\;c\ne c'; a\ne j\right\}$\end{itemize}

\item Data is removed from queues $U_{\sink_c}^{cv}$, $P_j^{c'cj}$, and $P_j^{\{ c,c'\}j}$.\end{itemize}

\subsection{Wireless case}

We also consider the multiple unicasts problem and corresponding modified problem in a wireless setting, using
the following wireless network model. We model wireless transmissions by generalized links, denoted by $(a,Z)$,
where $a$ is the originating node and $Z$ is the set of destination nodes. The network connectivity and link
transmission rates depend on the transmitted signal and interference powers according to some underlying
physical layer model. For example, the transmission rate per unit bandwidth $\mu_{ij}$ from node $i$ to node
$j$, with other nodes $n \in {\cal N}$ transmitting independent information simultaneously, may be given by the
Shannon formula \cite{cover:bk}
$$
\mu_{ij}(\underline{P}, \underline{S}) = \log \left( 1 + \frac{P_{i}S_{ij}}{N_{0}+\sum_{n \in {\cal
N}}P_{n}S_{nj}} \right)
$$
where $P_{l}$ is the power transmitted by node $l$, $S_{lj}$ is the channel gain from node $l$ to node $j$ and
$N_{0}$ is additive white Gaussian noise power over the signaling bandwidth. For simplicity, we consider a
finite set $\links$ of links and  a finite set $\scenarios$ of sets of simultaneously achievable link rates. We
denote by $\capacity_{(a,Z),\scenario}$ the capacity of link $(a,Z)$ in set $\scenario\in\scenarios$. A solution
to the multiple unicasts problem consists of a convex combination of sets in $\scenarios$, which gives a set of
average link capacities achievable by timesharing, and a network code that operates over the network with these
average link capacities.

In the wireless case, each node has a virtual source link, coding link and decoding link but no virtual
branching link; a branching operation occurs over a real wireless link. For virtual source, coding and decoding
links $e$, the set $\transmissions_{e}$ is defined exactly as in the wired case. For real wireless links
$(a,Z)$,
\begin{eqnarray*}\transmissions_{(a,Z)} &=&\{(U_a^{cv},U_b^{cv}),(U_a^{cv},U_b^{ca}),
(P_b^{\{c,c'\}j},P_a^{\{c,c'\}j}), (P_b^{cc'j},P_a^{cc'j}),(R_a^{cc'j},R_b^{cc'j}),\\&
&((U_a^{cv},R_a^{c'cj}),(U_b^{ca}, R_{b'}^{c'cj})) ,((P_{b}^{cc'j},P_{b'}^{c'cj}),P_a^{\{c,c'\}j}):\; c\ne
c';b,b'\in Z\}\end{eqnarray*}

\section{Back-pressure approximation algorithm}

Queues $U_a^{cv},U^{c},R_{a}^{cc'v},P_a^{cc'j}$ are associated with session $c$. We consider each joint poison
queue $P_j^{\{ c,c'\}j}$ as a pair of queues, one associated with each session $c$ and $c'$, such that equal
amounts are always added or removed from each of the pair of queues.

The overflow queue for each session $c$ has a potential $\parameter_c l e^{\parameter_c \maxQc}$ that is a
function of its length $l$, where \begin{eqnarray}\label{parameter}\parameter_c
=\fraction{\epsilon}{24Fr_c},\end{eqnarray}and $B$ is a constant to be determined. We divide each other queue
into subqueues, one for each link for which it is an origin or destination. Each subqueue $Q$ has a potential
$\potential_{c(Q)}(\length_Q) =e^{\parameter_{c(Q)} \length_Q}$ that depends on its length $\length_Q$ and its
session $c(Q)$. For notational simplicity, we will abbreviate subscripts $c(Q)$ as subscripts $Q$.

Flow entering or leaving subqueues associated with session $c$ is partitioned into packets of size
$\packetsize_c =(1+\epsilon)\inflow_c$. Besides its true length $\length_Q$, each subqueue $Q$ has an
approximate length $\approximate_Q$ that is an integer multiple of the packet size. The approximate length of a
subqueue is updated only when its true length has changed by at least one packet since the last update of its
approximate length, as follows: $\approximate_Q$ is set to $k\packetsize_Q$ where $k=\left\lfloor\fraction{
\length_Q-1}{ \packetsize_Q}\right\rfloor$ if $ Q$ is an origin subqueue or $k=\left\lceil\fraction{
\length_Q+1}{ \packetsize_Q}\right\rceil$ if $Q$ is a destination subqueue. Between updates, $\length_Q$ and
$\approximate_Q$ satisfy $\length_Q-3\packetsize_Q\less \approximate_Q\less \length_Q$ for an origin subqueue,
or $\length_Q\less \approximate_Q\less \length_Q+3\packetsize_Q$ for a destination subqueue.

We denote by $\bar{\transmissions}_e$ the subset of $\transmissions_e$ consisting of pairs
$(\origins,\destinations)\in\transmissions_e$ satisfying
\begin{eqnarray*}\min_{Q\in\origins}\approximate_Q&> &0\\
\max_{Q\in\destinations}\approximate_Q&<&Br_Q +\ln((L+1)\ratio)/\parameter_Q+3\packetsize_Q,\end{eqnarray*}where
$$\ratio=\fraction{\max_cr_c}{\min_cr_c}.  $$

\subsection{Wired case}

In each round $t$, the algorithm carries out the following:

\begin{enumerate}\item  Add $(1+\epsilon) \inflow_c$ units to
the overflow queue $\bU^c$ of each session $c$, then transfer as much as possible to $U^c$ subject to a maximum
length constraint of $\maxQc$ for $U^c$\item For each real and virtual link $e$, flow is pushed for zero or more
origin-destination pairs $(\origins,\destinations)\in\bar{\transmissions}_e$ such that the total amount pushed
is at most the link capacity $\capacity_e$. Specifically, initialize $\capacity$ to $\capacity_e$ and repeat
\begin{itemize}\item Choose the pair $(\origins,\destinations)\in\bar{\transmissions}_e$ that maximizes
\begin{equation}\decrease_{(\origins,\destinations)} =
\sum_{Q\in\origins}\potential_c'(\approximate_Q)-\sum_{Q\in\destinations}\potential_c'(\approximate_Q).
\label{decrease}\end{equation}Let
$$\capacity'=\min\left(\capacity,
\;\min_{Q\in(\origins\cup\destinations)}\left(\packetsize_Q-|\change_Q|\right)\right)$$
where $\change_Q$ is the change in $\length_Q$ since the last update of $\approximate_Q$.  Subtract $\capacity'$
units from $\capacity$, subtract $\capacity'$ units from $\length_Q$ for each $Q\in\origins$ and add units
$\capacity'$ to $\length_Q$ for each $Q\in\destinations$. For $Q\in(\origins\cup\destinations)$, if
$\capacity'=\packetsize_Q- |\change_Q|$, update $\approximate_Q$. \item If $\capacity  =0$ then
end.\end{itemize}\item Zero out all subqueues $U_{\sink_c}^{cv}$, $P_j^{c'cj}$ and $P_j^{\{ c,c'\}j}$. \item For
each queue that has at least one subqueue whose actual length has changed during the round, reallocate data
units to equalize the actual lengths of all its subqueues. If the actual length of any subqueue has changed by
at least one packet since the last update of its approximate length, update its approximate
length.
\end{enumerate}

When the amount of flow remaining in the network queues is an $\epsilon$-fraction of the total amount that has
entered the network, the flow values for each link are averaged over all rounds to give the solution.

\subsection{Wireless case}

The algorithm is the same as for the wired case, except for Phase 2, which is as follows.  For each virtual link
$e$, flow is pushed for zero or more origin-destination pairs
$(\origins,\destinations)\in\bar{\transmissions}_e$ exactly as in the wired case. For each real link $e=(a,Z)$,
flow is pushed for zero or more origin-destination pairs $(\origins,\destinations)\in\bar{\transmissions}_{e}$
such that the total amount pushed is at most the average link capacity
$\timeshare_{\scenario}\capacity_{e,\scenario}$ for some $\timeshare_{\scenario}$ satisfying
$\sum_{\scenario\in\scenarios}\timeshare_{\scenario}\less 1$.

Specifically, initialize $\duration$ to 1 and repeat
\begin{itemize}\item For each real link $e=(a,Z)$, let
\begin{eqnarray*}\decrease_{e} &=&\max_{(\origins,\destinations)\in\bar{\transmissions}_{ e}}
\decrease_{(\origins,\destinations)}\\(\origins_e,\destinations_e)
&=&\arg\max_{(\origins,\destinations)\in\bar{\transmissions}_{ e}}
\decrease_{(\origins,\destinations)}\end{eqnarray*} where $\decrease_{(\origins,\destinations)}$ is defined in
(\ref{decrease}). Choose the set $\scenario\in\scenarios$ which
maximizes\begin{equation}\label{score}\score_\scenario =\sum_{e} \decrease_{e}\capacity_{e,\scenario}.
\end{equation} Let
\begin{eqnarray*}\duration' =\min\left(\duration,\min_{e:\;\capacity_{e,\scenario}>0}
\min_{Q\in(\origins_e\cup\destinations_e)}\fraction{\packetsize_Q-|\change_Q|}{\capacity_{e,\scenario}}\right)
\end{eqnarray*}
where $\change_Q$ is the change in $\length_Q$ since the last update of $\approximate_Q$.  Subtract $\duration'$
units from $\duration$, and for each $e$, subtract $\duration'\capacity_{e,\scenario}$ units from $\length_Q$
for each $Q\in\origins_e$ and add the same amount to $\length_Q$ for each $Q\in\destinations_e$, updating
$\approximate_Q$ if  $\duration'\capacity_{e,\scenario}=\packetsize_Q- |\change_Q|$.\item If $\duration=0$ then
end.\end{itemize}

\section{Potential analysis}
Our analysis is partly based on the approach in~\cite{awerbuch94improved}. We lower bound the decrease in
potential over a round $t$. We denote by $Q(t)$ the actual length of a subqueue $Q$ at the end of Phase 1 of
round $t$. From~\cite{awerbuch94improved}, the increase in potential during Phase 1 is upper bounded by
\begin{equation}\sum_c\packetsize_c\potential_c'(U^{c}(t)). \label{phase1}\end{equation}

\subsection{Flow solution-based algorithm}
We lower bound the decrease in potential during Phases 2 and 3 by comparison with the potential decrease
resulting from pushing flow based on a flow solution for rates $f_c=(1+2\epsilon)\inflow_c$. This algorithm
differs from the back pressure algorithm only in the specific portion of Phase 2 determining the amount of flow
pushed for each pair $(\origins,\destinations)\in\bar{\transmissions}_e$ of each link $e$.

Consider a flow solution for rates $f_c$. Partition the flow for each session $c$ into elementary flows
$\flow_n^c$ such that all data in an elementary flow undergoes the same routing and coding operations. Each
elementary flow $\flow_n^c$ has a size denoted $\flowsize_n^c$, and consists of a set of links with associated
origin and destination subqueues, comprising a primary path from $\source_c$ to $\sink_c$ and a remedy path
associated with each coding node. Note that each flow $\flow_n^c$ starts from queue $U^c$, which consists of a
single subqueue, and that $\flowsize_c=\sum_n\flowsize_n^c$. Let $L$ be the length of the longest primary path
and $F$ the maximum number of links in an elementary flow. A subqueue in $\flow_n^c$ is considered upstream or
downstream of another according to the direction of flow in the modified problem defined in
Section~\ref{modified}.

Phase 2 of the flow solution-based algorithm consists of a preprocessing part and a flow pushing part.

\subsubsection{Preprocessing procedure}

\begin{itemize}\item Initialization:

Remove from each $\flow_n^c$ any portions of each branch that are downstream of a subqueue $Q$ in $\flow_n^c$
for which $\length_Q\less 3\packetsize_c$. Note that all subqueues of each queue have been equalized  in Phase 4
of the previous round. Let $\limiting$ be the set of flows $\flow^{c}_n$ containing some subqueue of length at
least $\maxQc +\ln ((L+1)\ratio)/\parameter_c$. For flows $\flow_n^c\in\limiting$, let $\bQ^{c}_n$ be the
furthest downstream origin subqueue in $\flow_n^c$ of length at least $\maxQc +\ln((L+1)\ratio)/\parameter_c$;
for flows $\flow_n^c\notin\limiting$, let $\bQ^{c}_n$ be the longest origin subqueue in $\flow_n^c$ (if there is
a tie, choose the furthest downstream).

Initialize $\flows$ as the set $\flows_0$ of all flows $\flow^{*c}_n$, initialize $\flows_1$ and $\flows_2$ as
empty sets, and for each $ c,n$, initialize $Q^{*c}_n$ as $\bQ^{c}_n$.

\item Phase A: Repeat\begin{itemize}\item Choose some flow $\flow^{c}_n\in\limiting\cap\flows$ and remove from
$\flows$ all flows $\flow^{c'}_{n'}$ such that $\flow^{c'}_{n'}$ shares any portion of a common joint poison
path segment, or the coding and branching links at either end, with $\flow^{c}_n$. Remove $\flow^c_n$ from
$\flows$ and add it to $\flows_1$.\item If there is no such flow remaining, end Phase A.\end{itemize}

\item Phase B: Associate with each flow $\flow^{c}_n\in\flows$ a weight $\weight^c_n$, initially set to
$\potential_c'(U^c(t))$.  
For each flow $\flow^{c'}_{n'}\in\flows$, set $\intersections^{c'}_{n'}$ to be the set of subqueues $\bQ^{c}_n$
such that $\flow^{c'}_{n'}$ is the shared flow corresponding to $\bQ^{c}_n$, and either
\begin{itemize}\item $\bQ^{c}_n$ is a remedy or individual poison subqueue whose corresponding branching link is in
$\flow^{c}_n$ and whose corresponding coding link is in $\flow^{c'}_{n'}$, or\item $\bQ^{c}_n$ is a joint poison
subqueue whose corresponding branching link is in $\flow^{c'}_{n'}$.\end{itemize}

Repeat\begin{itemize}\item

Choose some flow $\flow^{c'}_{n'}\in\flows$ such that
\begin{equation}\label{intersections}\weight^{c'}_{n'}\less\sum_{ \bQ^{c}_n\in\intersections^{c'}_{n'}}
\left(\potential_c'(\bQ^{c}_n(t))-\weight^{c}_n\right),\end{equation}\begin{itemize}\item remove
$\flow^{c'}_{n'}$ from $\flows$, \item remove any subqueues in $\flow^{c'}_{n'}$ from
$\intersections^c_n\;\forall\; c,n$, \item set the weight $\weight^{ c}_n$ of each flow
$\flow^{c}_n\in\intersections^{c'}_{n'}$ to $\potential_c'(\bQ^{c}_{n}(t))$.\end{itemize}

\item If there is no such flow, for each $\intersections^{c'}_{n'}$ and each $
Q^{*c}_{n}\in\intersections^{c'}_{n'}$ set $Q^{*c}_{n}$ to $U^c$, and end Phase B.
\end{itemize}
\item Phase C: Repeat\begin{itemize}\item Choose some flow $\flow^{c}_n\in\flows$ such that the longest
individual poison subqueue along one of its poison branches, $Q$, is longer than $Q^{*c}_n$, and the entire
joint portion of that branch together with the branching link are not in the corresponding shared flow.  Set
$Q^{*c}_{n}$ to $Q$ and remove from $\flow^c_n$ all but the portion downstream of $Q$. \item If there is no such
flow remaining, end Phase C.\end{itemize}

\end{itemize}

Observe that:\begin{itemize} \item At most $L$ flows are removed by each flow in $\limiting$,
and\begin{eqnarray}\nonumber\potential_c'(\maxQc
+\ln((L+1)\ratio)/\parameter_c)&\greater&(L+1)\max_c\potential_c'(\maxQc)\\&\greater&(L+1)\max_c\potential_c'(
U^c(t)\label{phaseA}\end{eqnarray}\item $\flowsize^c_n=\flowsize^{c'}_{n'}$ for all $\flow^{c}_n$ and
$\flow^{c'}_{n'}$ that share a common joint poison path segment.\item At the end of the preprocessing procedure,
\begin{equation}\label{invariant2}2\sum_{\flow^{c}_{n}\in\flows}\potential_c'(Q^{*c}_{n}(t))
\flowsize^c_n \greater\sum_{ \flow^{c}_{n} \in\flows} \potential_c'(\bQ^c_n(t)) \flowsize^c_n.\end{equation}To
see this, note that at the end of Phase B there is a one-to-one correspondence between flows
$\flow^{c}_n\in\flows$ for which $Q^{*c}_n\ne\bQ^c_n$ and elements $\bQ^c_n$ in the sets
$\intersections^{c'}_{n'}$ of flows $\flow^{c'}_{n'}\in\flows$.  For each $\flow^{c'}_{n'}\in\flows$,  by
(\ref{intersections}),
$$\weight^{c'}_{n'}>\sum_{ \bQ^{c}_n\in\intersections^{c'}_{n'}}
\left(\potential_c'(\bQ^{c}_n(t))-\weight^{c}_n\right).
$$Multiplying by $\flowsize^{c}_n$ and summing over all $\flow^{c'}_{n'}\in\flows$, and noting that at the end
of the preprocessing procedure
\begin{equation}\label{weight}\potential_c'(Q^{*c}_{n}(t))\greater \weight^{c}_{n}\end{equation} gives
(\ref{invariant2}).\item Phase B and C maintain the invariant
\begin{equation}\label{invariant1}\sum_{\flow^{c}_{n}\in\flows}\weight^c_n\flowsize^c_n \greater\sum_{ \flow^{c}_{n}
\in\flows_2} \potential_c'(U^c(t)) \flowsize^c_n\end{equation}where $\flows_2$ is the value of $\flows$ at the
start of Phase B. The invariant holds since both sides are equal at the start of Phase B, and the left-hand side
is monotonically non-decreasing. From (\ref{phaseA}), (\ref{weight}) and (\ref{invariant1}), at the end of the
preprocessing procedure we
have\begin{equation}\label{input-coverage}\sum_{\flow^{c}_n\in(\flows_1\cup\flows)}\potential_c'(Q^{*c}_{n}(t))\flowsize^c_n
\greater\sum_{ \flow^{c}_{n}\in\flows_0}\potential_c'(U^c(t)) \flowsize^c_n.\end{equation}\end{itemize}

\subsubsection{Flow pushing procedure}

For each flow $\flow^{c}_n\in\flows_1\cup\flows$, let $\flow^{*c}_n$ be the portion of $\flow_n^c$ downstream of
$Q^{*c}_n$. Partition its links into a set $\subsets^{*c}_n$ of subsets. Each of these subsets
$\subflow\in\subsets^{*c}_n$ may be
\begin{itemize}\item a path from $Q^{*c}_n$, if it is a poison subqueue, to its associated coding node
$a(\subflow)$ \item a path from $Q^{*c}_n$, if it is a remedy subqueue, up to and including its associated
decoding link at node $w(\subflow)$, and the associated poison path from $w(\subflow)$ to its associated coding
node $a(\subflow)$ via the branching link at node $b(\subflow)$,\item a path associated with uncoded flow ending
in sink node $\sink_c$, or\item a path associated with uncoded flow ending in a coding link at a node
$a(\subflow)$, together with the associated remedy path up to and including the decoding link at a node
$w(\subflow)$, and the associated poison path from $w$ to $a$ via the branching link at a node
$b$.\end{itemize}Note that each subset starts either at $Q^{*c}_n$ or at an uncoded subqueue.

The flow solution-based algorithm pushes flow as follows. First, for each pair $(\subflow_c
,\subflow_{c'})\in\subsets^{*c}_n\times\subsets^{*c'}_{n'}$ that shares any portion of a common joint poison
path segment, note that $\flowsize_n^c=\flowsize^{c'}_{n'}$ and that Phase B of the preprocessing procedure
ensures that both subsets or neither contains the coding link; in the latter case both or neither contains the
branching link.  Thus, the following are the only two cases:\begin{itemize} \item Case 1: One of the subsets,
say $\subflow_c$, contains both the coding link at $a(\subflow_c)$ and the branching link at $b(\subflow_c)$,
while $\subflow_{c'}$ contains the coding link but not the branching link. Phase C of the preprocessing
procedure ensures that all session $c$ individual poison subqueues along the joint poison path segment are
shorter than $Q^{*c}_n$. Push $\flowsize_n^c$ units through $\subflow_c \cup\subflow_{c'}$, pushing session $c$
individual poison units through the joint poison path segment.\item Case 2: Both subsets contain the same
portion of the joint poison path segment. Push $\flowsize_n^c$ units through $\subflow_c
\cup\subflow_{c'}$.\end{itemize}

Next, for each subset $\subflow$ of some $\flow^{*c}_n$ that does not have any coded segments in common with
$\flow^{*c'}_{n'}$ for all $c'\ne c,n'$, we have the following cases:\begin{itemize}\item Case 1: $Q^{*c}_n$ is
an individual poison subqueue in $\subflow$. Phase B of the preprocessing procedure ensures that $Q^{*c}_n$ is
the longest individual poison subqueue in $\flow^{*c'}_{n'}$. Push $\flowsize^{c}_n$ individual poison units
through $\subflow$. \item Case 2: $Q^{*c}_n$ is a joint poison subqueue in $\subflow$. Push $\flowsize^{c}_n$
joint poison units along the path in $\subflow$ downstream of $Q^{*c}_n$. \item Case 3: $Q^{*c}_n$ is a remedy
subqueue in $\subflow$. Phase C of the preprocessing procedure ensures that $Q^{*c}_n$ is longer than all
session $c$ individual poison subqueues along the primary path of $\subflow$.  Push $ \flowsize^{c}_n$ remedy
units along $\subflow$ through the decoding link at $w(\subflow)$, and $\flowsize^{c}_n$ individual poison units
along the primary path of $\subflow$.\item Case 4: $Q^{*c}_n$ is an uncoded subqueue or is not in $\subflow$.
Push $\flowsize^{c}_n$ uncoded units along the primary path of $\subflow$ starting from its longest session $c$
uncoded subqueue.
\end{itemize}

Note that flow is pushed only from origin subqueues $Q$ for which
$$\length_Q>3\packetsize_Q\Rightarrow\approximate_Q>0$$and only to destination subqueues $Q$ for which $$
\length_Q<\maxQQ +\ln ((L+1)\rho)/\parameter_Q\Rightarrow\approximate_Q<\maxQQ +\ln
((L+1)\rho)/\parameter_Q+3\packetsize_Q.  $$

\subsection{Potential decrease in Phases 2 and 3}
The decrease in potential from pushing $f$ units across a link from a set $\origins$ of origin to a set
$\destinations$ of destination subqueues is at least
$$\sum_{Q\in\origins}\left(f\potential_Q'(\length_Q)-f^2\potential_Q''(\length_Q)\right)-
\sum_{Q\in\destinations}\left(f\potential_Q'(\length_Q) + f^2\potential_Q''(\length_Q+f)\right) $$where
$\length_Q$  denotes the initial length of each subqueue $Q$.

\subsubsection{Flow solution-based algorithm}

We denote by $\origins(\flow^{*c}_n)$ and $\destinations(\flow^{*c}_n)$ the sets of origin and destination
subqueues of a flow $\flow^{*c}_n$, and by $\origins_{ c,e}$ and $\destinations_{ c,e}$ the sets of session $c$
origin and destination subqueues of a link $e$. For each subqueue $Q$, denote by $\flowsize_Q$ the total flow
out of $Q$ (if $Q$ is an origin subqueue) or into $Q$ (if $Q$ is a destination subqueue) in the flow pushing
procedure of the flow-solution based algorithm.
The potential drop over Phases 2 and 3 in the flow solution-based algorithm is at
least\begin{eqnarray*}\nonumber & &\hspace{- 0.2 in} \sum_{ c,e}\left(\sum_{Q\in\origins_{
c,e}}\left(\flowsize_Q\potential_c'(\length_Q)-\flowsize_Q^2 \potential_c''(\length_Q+\flowsize_Q)\right)-
\sum_{Q\in\destinations_{ c,e}}\flowsize_Q \left(\potential_c'(\length_Q) +\flowsize_Q^2
\potential_c''(\length_Q+\flowsize_Q)\right)\right)\\& \greater& \sum_{c,e}\left(\sum_{Q\in\origins_{
c,e}}\flowsize_Q \left(\potential_c'(\length_Q)-\flowsize_c\potential_c''(\length_Q+\flowsize_c)\right)-
\sum_{Q\in\destinations_{ c,e}}\flowsize_Q \left(\potential_c'(\length_Q)
+\flowsize_c\potential_c''(\length_Q+\flowsize_c)\right)\right)\\& \greater&
\sum_{\flow^{*c}_n}\left(\sum_{Q\in\origins(\flow^{*c}_n)}\flowsize^c_n
\left(\potential_c'(\length_Q)-\flowsize_c\potential_c''(\length_Q+\flowsize_c)\right)-
\sum_{Q\in\destinations(\flow^{*c}_n)}\flowsize^c_n\left(\potential_c'(\length_Q)
+\flowsize_c\potential_c''(\length_Q+\flowsize_c)\right)\right)\end{eqnarray*} Since all subqueues of each queue
have been equalized in Phase 4 of the previous round, and since each flow $\flow^{*c}_n$ has one origin subqueue
of length $Q^{*c}_n(t)$, at most $L +1$ destination subqueues each of length less than $3\packetsize_c$, and a
total of at most $F$ links, this potential drop is lower-bounded by
\begin{eqnarray} & &\hspace{- 0.55in }\sum_{
\flow^{c}_n\in(\flows_1\cup\flows)}\hspace{- 0.23in}
\flowsize_n^c\left(\potential_c'(Q^{*c}_n(t))-\left(L+1\right) \potential_c'(3\packetsize_c)\right)-\hspace{-
0.125in } \sum_{\flow^{c}_n\in\flows_1}\hspace{- 0.15 in}3F\flowsize_n^c\flowsize_c\potential_c''(Q^{*c}_n(t)+
\flowsize_c)-\sum_{\flow^{c}_n\in\flows} 3F\flowsize_n^c\flowsize_c\potential_c''(\bQ^{c}_n(t)+
\flowsize_c)\nonumber\end{eqnarray}

From (\ref{parameter}), we have\begin{eqnarray}3F\flowsize_c\potential_c''(l+\flowsize_c)&\less &
3F(1+2\epsilon)\inflow_c\potential_c''(l+(1+2\epsilon)\inflow_c)\nonumber\\& =
&3F(1+2\epsilon)\inflow_c\parameter_c^2 e^{\parameter_c l+\parameter_c(1+2\epsilon)\inflow_c}\nonumber
\\& = & 3F(1+2\epsilon)\inflow_c\parameter_c e^{\parameter_c(1+2\epsilon)\inflow_c}\potential_c'(l)\nonumber
\\& = & \fraction{\epsilon(1+2\epsilon)}{8}e^{\fraction{\epsilon(1+2\epsilon)}{24F}}\potential_c'(l)\nonumber
\\&\less
&\epsilon\potential_c'(l)/4. \label{derivative2}\end{eqnarray} This yields, using (\ref{invariant2}), the
following lower bound on the potential drop:\begin{eqnarray} &&\hspace{- 0.3
in}\sum_{\flow^{c}_n\in(\flows_1\cup\flows)}\hspace{- 0.23in
}\flowsize_n^c\left(\potential_c'(Q^{*c}_n(t))-\left(L+1\right) \potential_c'(3\packetsize_c)\right)-\hspace{-
0.125in } \sum_{\flow^{c}_n\in\flows_1}\hspace{- 0.15in}
\flowsize_n^c\epsilon\potential_c'(Q^{*c}_n(t))/4-\sum_{\flow^{c}_n\in\flows}\flowsize_n^c\epsilon\potential_c'(\bQ^{c}_n(t))/4\nonumber\\&\greater
&\sum_{\flow^{c}_n\in(\flows_1\cup\flows)}
\flowsize_n^c\left(\left(1-\fraction{\epsilon}{2}\right)\potential_c'(Q^{*c}_n(t))-\left(L+1\right)
\potential_c'(3\packetsize_c)\right) \nonumber\\&\greater &\sum_c
(1+2\epsilon)\inflow_c\left(\left(1-\fraction{\epsilon}{2}\right)\potential_c'(U^c(t))-\left(L
+1\right)\potential_c'(3\packetsize_c)\right) \nonumber\\&\greater &\sum_c
\inflow_c\left(\left(1+\fraction{3\epsilon}{2}-\epsilon^2\right)\potential_c'(U^c(t))-\left(L
+1\right)(1+2\epsilon)\potential_c'(3\packetsize_c)\right) \label{potential-drop}\end{eqnarray}

\subsubsection{Back pressure algorithm}
If the packet size $\packetsize_c$ were infinitesimally small rather than $(1+\epsilon)\inflow_c$, then
$\length_Q=\approximate_Q$ and the procedure in Phase 2 of the back pressure algorithm would give, for each link
$e$, the maximum possible potential decrease from pushing flow for zero or more origin-destination pairs
$(\origins,\destinations)\in\bar{\transmissions}_e$ such that the total amount pushed is at most the link
capacity $\capacity_e$. Since $\packetsize_c=\Theta(\inflow_c)$ and
$$\potential_c'(l+\Theta(\inflow_c)) -\potential_c'(l)\less\Theta(\inflow_c)\potential_c''(l+\Theta(\inflow_c)),
$$the back pressure algorithm achieves, for each link $e$, a potential decrease of at
least that achieved by the flow solution-based algorithm minus an error
term$$\sum_{Q\in(\origins_e\cup\destinations_e)}\Theta(f_Q
\inflow_Q)\potential_Q''(\length_Q+\Theta(\inflow_Q))$$where $f_Q$ is the amount of flow added or removed from
$Q$ in Phase 2 of the flow solution-based algorithm. Since $f_Q=\Theta(\inflow_Q)$, decreasing each
$\parameter_c$ by some constant factor is sufficient to ensure that (\ref{potential-drop}) applies to the back
pressure algorithm.

\subsection{Overall potential change  and number of rounds}
The potential does not increase during  Phase 4.  Thus, from (\ref{phase1}) and (\ref{potential-drop}), the
overall potential decrease during the round is lower bounded by \begin{eqnarray*} & &\hspace{- 0.2 in}\sum_c
\inflow_c\left(\left(\fraction{\epsilon}{2}-\epsilon^2\right)\potential_c'(U^c(t))-\left(L
+1\right)(1+2\epsilon)\potential_c'(3\packetsize_c)\right)\\& = &\sum_c
\inflow_c\left(\fraction{\epsilon}{2}-\epsilon^2\right)\potential_c'(U^c(t))-\left(L
+1\right)(1+2\epsilon)\fraction{\epsilon}{24F}e^{\fraction{\epsilon(1+\epsilon)}{8F}}\end{eqnarray*}If
$U^c(t)=\maxQc $ for some $c$, then the decrease in potential is at least \begin{eqnarray*} & &\hspace{- 0.1 in}
\inflow_c\left(\fraction{\epsilon}{2}-\epsilon^2\right)\potential_c'(\maxQc ) -K\left(L
+1\right)(1+2\epsilon)\fraction{\epsilon}{24F}e^{\fraction{\epsilon(1+\epsilon)}{8F}}\\& = &
\inflow_c\parameter_c\left(\fraction{\epsilon}{2}-\epsilon^2\right)e^{\parameter_c \maxQc } -K\left(L
+1\right)(1+2\epsilon)\fraction{\epsilon}{24F}e^{\fraction{\epsilon(1+\epsilon)}{8F}}\end{eqnarray*}which is
non-negative if
\begin{eqnarray*}\maxQc &=&\fraction{1}{\parameter_c}\ln\left(\fraction{K\left(L
+1\right)(1+2\epsilon)}{\epsilon(1-2\epsilon)}\right) +3\packetsize_c\\& = &
\Theta\left(\fraction{1}{\parameter_c}
\ln\left( \fraction{KL}{\epsilon}\right)\right)\end{eqnarray*} If $U^c(t)<\maxQc $ for all $c$, the overflow
queues are empty and, since there are $\Theta(NMK)$ session-$c$ subqueues each of potential less than
$$\potential_c(\maxQc +\ln ((L+1)\rho)/\parameter_c +4\packetsize_c) =\Theta\left(L\rho e^{\fraction{\epsilon B}{24
F}}\right), $$the overall potential in the system at the end of the round is at most
$$\Theta\left(NMK^2L\rho e^{\fraction{\epsilon B}{24 F}}\right)$$
By induction, this is also an upper bound on the total potential at the end of every round. The length of the
overflow queue for session $c$ is thus never more than
$$O\left(\fraction{NMK^2L\ratio e^{\fraction{\epsilon B}{24 F}}}{\parameter_c e^{\fraction{\epsilon B}{24 F}}}\right)=
O\left(\fraction{NMK^2L\ratio}{\parameter_c}\right)
$$
and the total units for session $c$ is at most \begin{eqnarray*}O\left(\fraction{NMK^2L\ratio }{\parameter_c}
+NMK(\maxQc +\ln (L\rho)/\parameter_c +4\packetsize_c)\right) & = &O\left(\fraction{NMK }{\parameter_c}
\left(KL\ratio+ \ln\left(\fraction{KL\ratio}{\epsilon}\right)\right)\right)
\\& = &O\left(\fraction{NMK F\inflow_c}{\epsilon}\left(KL\ratio+
\ln\left(\fraction{KL\ratio}{\epsilon}\right)\right)\right)
\end{eqnarray*}

Thus, at most $$O\left(\fraction{NMKF }{\epsilon} \left(KL\ratio+
\ln\left(\fraction{KL\ratio}{\epsilon}\right)\right)\right)$$rounds of input flow for each session remain in the
network at any time.  For $$t=O\left(\fraction{NMKF }{\epsilon^2} \left(KL\ratio+
\ln\left(\fraction{KL\ratio}{\epsilon}\right)\right)\right)$$ amount remaining in the network is at most a
fraction $\epsilon$ of the total amount that has entered the network up to  round $t$.

\subsection{Number of operations}

In each round, at each node, pushing flow across links results in a total decrease of at most $O(\bcapacity)$ in
the actual lengths of origin subqueues and a total increase of at most $O(\bcapacity)$ in the actual lengths of
destination subqueues. The total change in subqueue lengths from rebalancing is not more than the total change
resulting from pushing flow across links. Thus, at most $ O(ND)$ approximate subqueue lengths in the network are
updated in each round, where $D=\max_c\bcapacity/\inflow_c$. Assuming the subqueue differences computed by the
algorithm can be stored, only those differences involving subqueues whose approximate lengths have changed are
recomputed.

For the coding link at each node $i$, for each pair of sessions $\{c,c'\}$, the values of
$\decrease_{(\origins,\destinations)}$, defined in (\ref{decrease}), for pairs
$(\origins,\destinations)\in\bar{\transmissions}_e$ of the form
$((U_i^{cv},U_i^{c'v'}),(R_{v'}^{cc'i},R_v^{c'ci}))$ are stored in a sorted list of length $O(N^2)$, and the
list indexes are ranked according to the maximum value in each list. If $U_i^{cv}$ or $U_i^{c'v'}$ changes, $N$
differences are updated in each of $K$ lists, requiring $O(NK\log N)$ operations, and the rank of these $K$
lists are updated using $O(K\log K)$ operations. If $R_{v'}^{cc'i}$ or $R_v^{c'ci}$ changes, $N$ differences are
updated in one of the lists and fewer operations are required.

For the decoding link $e$ at each node $i$, the values of $\decrease_{(\origins,\destinations)}$ for pairs
$(\origins,\destinations)\in\bar{\transmissions}_e$
are stored in a sorted list of length $O(NK^2)$. A change to any of the approximate subqueue lengths requires
updating of at most $K$ differences, for which $O(K\log(NK))$ operations suffices.

\subsubsection{Wired case}

For each real and virtual branching link, the values of $\decrease_{(\origins,\destinations)}$ for pairs
$(\origins,\destinations)\in\bar{\transmissions}_e$ are similarly stored in a sorted list of length $O(NK^2)$. A
change to any of the approximate subqueue lengths requires updating of $O(1)$ differences, for which
$O(\log(NK))$ operations suffices.

Thus, each round has complexity $O(N^2KD\log (NK))$, and the algorithm has complexity
$$O\left(\fraction{N^3MK^2FD\log (NK) }{\epsilon^2} \left(KL\ratio+
\ln\left(\fraction{KL\ratio}{\epsilon}\right)\right)\right).$$

\subsubsection{Wireless case}
We assume that for each wireless link $(a,Z)$, $|Z|=O(1)$. For a wireless link $e=(a,Z)$, for each pair of
sessions $\{c,c'\}$, the values of $\decrease_{(\origins,\destinations)}$ for pairs
$(\origins,\destinations)\in\bar{\transmissions}_e$ of the form
$((U_a^{cv},R_{a}^{c'cj}),(U_b^{ca},R_{b'}^{c'cj}))$ are stored in a sorted list of length $O(N^2)$, and the
values of $\decrease_{(\origins,\destinations)}$ for all other  pairs
$(\origins,\destinations)\in\bar{\transmissions}_e$ are stored in a sorted list of length $O(NK^2)$. The values
of $\score_\scenario$, defined in (\ref{score}), for each set $\scenario\in\scenarios$ are also stored.

Of all the subqueues of $(a,Z)$, a change in the approximate length of some subqueue $U_a^{cv}$ or $U_b^{ca}$
requires the most operations: $O(N)$ differences are updated in each of $O(K)$ length-$O(N^2)$ lists, $O(1)$
differences are updated in the length-$O(NK^2)$ list and $O(K)$ of the $O(K^2)$ list indexes are repositioned,
requiring a total of $O(NK\log N)+ O(\log(NK))+O(K\log K) = O(NK\log (NK))$ operations. A further
$O(|\scenarios|)$ operations suffices to update the values of $\score_\scenario$ and find the maximum among
them.

Thus, each round has complexity $O(ND(NK\log (NK)+|\scenarios|)$, and the algorithm has complexity
$$O\left(\fraction{N^2MKFD(NK\log (NK ) +|\scenarios|)}{\epsilon^2} \left(KL\ratio+
\ln\left(\fraction{KL\ratio}{\epsilon}\right)\right)\right).$$
\bibliographystyle{plain}
\bibliography{NWC-trace}
\end{document}